\documentstyle[twoside,epsfig]{article}
\catcode`\@=11
\long\def\@makefntext#1{
\protect\noindent \hbox to 3.2pt {\hskip-.9pt
$^{{\eightrm\@thefnmark}}$\hfil}#1\hfill}       

\def\thefootnote{\fnsymbol{footnote}}
\def\@makefnmark{\hbox to 0pt{$^{\@thefnmark}$\hss}}    
\def\ps@myheadings{\let\@mkboth\@gobbletwo
\def\@oddhead{\hbox{}
\rightmark\hfil\eightrm\thepage}
\def\@oddfoot{}\def\@evenhead{\eightrm\thepage\hfil
\leftmark\hbox{}}\def\@evenfoot{}
\def\sectionmark##1{}\def\subsectionmark##1{}}

\oddsidemargin=\evensidemargin
\renewcommand{\thefootnote}{\fnsymbol{footnote}}

\newcounter{sectionc}\newcounter{subsectionc}\newcounter{subsubsectionc}
\renewcommand{\section}[1] {\vspace{12pt}\addtocounter{sectionc}{1}
\setcounter{subsectionc}{0}\setcounter{subsubsectionc}{0}\noindent
    {\tenbf\thesectionc. #1}\par\vspace{5pt}}
\renewcommand{\subsection}[1] {\vspace{12pt}\addtocounter{subsectionc}{1}
    \setcounter{subsubsectionc}{0}\noindent
    {\bf\thesectionc.\thesubsectionc. {\kern1pt \bfit #1}}\par\vspace{5pt}}
\renewcommand{\subsubsection}[1] {\vspace{12pt}\addtocounter{subsubsectionc}{1}
    \noindent{\tenrm\thesectionc.\thesubsectionc.\thesubsubsectionc.
    {\kern1pt \tenit #1}}\par\vspace{5pt}}
\newcommand{\nonumsection}[1] {\vspace{12pt}\noindent{\tenbf #1}
    \par\vspace{5pt}}
\newcounter{appendixc}
\newcounter{subappendixc}[appendixc]
\newcounter{subsubappendixc}[subappendixc]
\renewcommand{\thesubappendixc}{\Alph{appendixc}.\arabic{subappendixc}}
\renewcommand{\thesubsubappendixc}
    {\Alph{appendixc}.\arabic{subappendixc}.\arabic{subsubappendixc}}
\renewcommand{\appendix}[1] {\vspace{12pt}
        \refstepcounter{appendixc}
        \setcounter{figure}{0}
        \setcounter{table}{0}
        \setcounter{lemma}{0}
        \setcounter{theorem}{0}
        \setcounter{corollary}{0}
        \setcounter{definition}{0}
        \setcounter{equation}{0}
        \renewcommand{\thefigure}{\Alph{appendixc}.\arabic{figure}}
        \renewcommand{\thetable}{\Alph{appendixc}.\arabic{table}}
        \renewcommand{\theappendixc}{\Alph{appendixc}}
        \renewcommand{\thelemma}{\Alph{appendixc}.\arabic{lemma}}
        \renewcommand{\thetheorem}{\Alph{appendixc}.\arabic{theorem}}
        \renewcommand{\thedefinition}{\Alph{appendixc}.\arabic{definition}}
        \renewcommand{\thecorollary}{\Alph{appendixc}.\arabic{corollary}}
        \noindent{\tenbf Appendix \theappendixc #1}\par\vspace{5pt}}
\newcommand{\subappendix}[1] {\vspace{12pt}
        \refstepcounter{subappendixc}
        \noindent{\bf Appendix \thesubappendixc. {\kern1pt \bfit #1}}
    \par\vspace{5pt}}
\newcommand{\subsubappendix}[1] {\vspace{12pt}
        \refstepcounter{subsubappendixc}
        \noindent{\rm Appendix \thesubsubappendixc. {\kern1pt \tenit #1}}
    \par\vspace{5pt}}
\newcommand{\textlineskip}{\baselineskip=13pt}
\newcommand{\smalllineskip}{\baselineskip=10pt}
\def\eightcirc{
\begin{picture}(0,0)
\put(4.4,1.8){\circle{6.5}}
\end{picture}}
\def\eightcopyright{\eightcirc\kern2.7pt\hbox{\eightrm c}}
\newcommand{\copyrightheading}[1]
    {\vspace*{-2.5cm}\smalllineskip{\flushleft
    {\footnotesize International Journal of Modern Physics C #1}\\
    {\footnotesize $\eightcopyright$\, World Scientific Publishing
     Company}\\
     }}

\newcommand{\publisher}[2]{{\begin{center}\footnotesize\smalllineskip
    Received #1\\
    Revised #2
    \end{center}
    }}

\def\abstracts#1#2#3{{
    \centering{\begin{minipage}{4.5in}\footnotesize\baselineskip=10pt
    \parindent=0pt #1\par
    \parindent=15pt #2\par
    \parindent=15pt #3
    \end{minipage}}\par}}

\def\keywords#1{{
    \centering{\begin{minipage}{4.5in}\footnotesize\baselineskip=10pt
    {\footnotesize\it Keywords}\/: #1
    \end{minipage}}\par}}

\renewenvironment{thebibliography}[1]
        {\frenchspacing
     \ninerm\baselineskip=11pt
         \begin{list}{\arabic{enumi}.}
        {\usecounter{enumi}\setlength{\parsep}{0pt}
     \setlength{\leftmargin 12.7pt}{\rightmargin 0pt} 
         \setlength{\itemsep}{0pt} \settowidth
    {\labelwidth}{#1.}\sloppy}}{\end{list}}

\newcounter{itemlistc}
\newcounter{romanlistc}
\newcounter{alphlistc}
\newcounter{arabiclistc}

\newcommand{\fcaption}[1]{
        \refstepcounter{figure}
        \setbox\@tempboxa = \hbox{\footnotesize Fig.~\thefigure. #1}
        \ifdim \wd\@tempboxa > 5in
           {\begin{center}
        \parbox{5in}{\footnotesize\smalllineskip Fig.~\thefigure. #1}
            \end{center}}
        \else
             {\begin{center}
             {\footnotesize Fig.~\thefigure. #1}
              \end{center}}
        \fi}
\newcommand{\tcaption}[1]{
        \refstepcounter{table}
        \setbox\@tempboxa = \hbox{\footnotesize Table~\thetable. #1}
        \ifdim \wd\@tempboxa > 5in
           {\begin{center}
        \parbox{5in}{\footnotesize\smalllineskip Table~\thetable. #1}
            \end{center}}
        \else
             {\begin{center}
             {\footnotesize Table~\thetable. #1}
              \end{center}}
        \fi}

\def\@citex[#1]#2{\if@filesw\immediate\write\@auxout
    {\string\citation{#2}}\fi
\def\@citea{}\@cite{\@for\@citeb:=#2\do
    {\@citea\def\@citea{,}\@ifundefined
    {b@\@citeb}{{\bf ?}\@warning
    {Citation `\@citeb' on page \thepage \space undefined}}
    {\csname b@\@citeb\endcsname}}}{#1}}
\newif\if@cghi
\def\cite{\@cghitrue\@ifnextchar [{\@tempswatrue
    \@citex}{\@tempswafalse\@citex[]}}
\def\citelow{\@cghifalse\@ifnextchar [{\@tempswatrue
    \@citex}{\@tempswafalse\@citex[]}}
\def\@cite#1#2{{$\null^{#1}$\if@tempswa\typeout
    {IJCGA warning: optional citation argument
    ignored: `#2'} \fi}}

\def\pmb#1{\setbox0=\hbox{#1}
    \kern-.025em\copy0\kern-\wd0
    \kern.05em\copy0\kern-\wd0
    \kern-.025em\raise.0433em\box0}

\def\fnt#1#2{\footnotetext{\kern-.3em
    {$^{\mbox{\scriptsize #1}}$}{#2}}}

\def\ps@myheadings{%
    \let\@oddfoot\@empty\let\@evenfoot\@empty
    \def\@evenhead{\slshape\leftmark\hfil}
    \def\@oddhead{\hfil{\slshape\rightmark}}
    \let\@mkboth\@gobbletwo
    \let\sectionmark\@gobble
    \let\subsectionmark\@gobble
    }

\font\tenrm=cmr10
\font\tenit=cmti10
\font\tenbf=cmbx10
\font\bfit=cmbxti10 at 10pt
\font\ninerm=cmr9

\font\eightrm=cmr8

\def\qed{\hbox{${\vcenter{\vbox{            
   \hrule height 0.4pt\hbox{\vrule width 0.4pt height 6pt
   \kern5pt\vrule width 0.4pt}\hrule height 0.4pt}}}$}}

\renewcommand{\thefootnote}{\fnsymbol{footnote}}    

\def\bsc{{\sc a\kern-6.4pt\sc a\kern-6.4pt\sc a}}   
\def\bflatex{\bf L\kern-.30em\raise.3ex\hbox{\bsc}\kern-.14em
T\kern-.1667em\lower.7ex\hbox{E}\kern-.125em X}

\begin{document}
\setlength{\textheight}{7.7truein}  
\thispagestyle{empty}
\markboth{\protect{\footnotesize\it Traffic Flow}}{\protect{\footnotesize\it Traffic Flow}}
\normalsize\textlineskip
\setcounter{page}{1}
\copyrightheading{}         
\vspace*{0.88truein}
\centerline{\bf A CELLULAR AUTOMATON MODEL FOR}
\centerline{\bf THE TRAFFIC FLOW IN BOGOT\'A}
\vspace*{0.035truein} \vspace*{0.37truein}
\centerline{\footnotesize L. E. Olmos} \baselineskip=12pt
\centerline{\footnotesize\it Dpto. de F\'{\i}sica, Univ. Nacional
de Colombia} \baselineskip=10pt \centerline{\footnotesize\it
Bogot\'{a} D.C, Colombia} \centerline{\footnotesize\it E-mail:
leolmoss@unal.edu.co}

\vspace*{15pt}          
\centerline{\footnotesize J.D. Mu\~noz} \baselineskip=12pt
\centerline{\footnotesize\it Dpto. de F\'{\i}sica, Univ. Nacional
de Colombia} \baselineskip=10pt \centerline{\footnotesize\it
Bogot\'{a} D.C., Colombia} \centerline{\footnotesize\it E-mail:
jdmunoz@ica1.uni-stuttgart.de} \vspace*{0.225truein}
\publisher{(received date)}{(revised date)}
\vspace*{0.25truein} \abstracts{In this work we propose a car cellular
  automaton model that reproduces the experimental behavior of traffic flows
  in Bogot\'a. Our model includes three elements: hysteresis between the
acceleration and brake gaps, a delay time in the acceleration, and an
instantaneous brake. The parameters of our model were obtained from direct
  measurements inside a car on motorways in Bogot\'a. Next, we simulated with this
  model the flux-density fundamental diagram for a single-lane traffic road
  and compared it with experimental data. Our simulations are
  in very good agreement with the experimental measurements, not just in the
  shape of the fundamental diagram, but also in the numerical values for both
  the road capacity and the density of maximal flux. Our model reproduces,
  too, the qualitative behavior of shock waves. In addition, our work
  identifies the periodic boundary conditions as the source of false peaks in
  the fundamental diagram, when short roads are simulated that have been also
  found in previous works. The phase 
  transition between free and congested traffic is also investigated by
  computing both the relaxation time and the order parameter. Our work shows
  how different the traffic behavior from one city to another can be, and how
  important is to determine the model parameters for each city.}{}{}

\vspace*{5pt} \keywords{Cellular automaton models, Fundamental diagrams, Jams, Shock waves, jamming transitions.}



\vspace*{1pt}\textlineskip \setcounter{footnote}{0}
\renewcommand{\thefootnote}{\alph{footnote}}
\section{Introduction}

Some years ago, Bogot\'a was a city with heavy traffic congestion and a
chaotic transportation system, just because it has 7 million
inhabitants with more than 55,000 taxis, 18,000 buses of different
kinds, and a million of private cars roaming the streets.  Recent
city administrations have tried to solve this problem by introducing
transportation strategies such as: a mass transportation system 
({\it Transmilenio}), almost 250 kilometres of bike paths, pedestrian bridges
everywhere and restrictions on the use of private cars at rush hours ({\it
  Pico y placa}). Thanks these efforts, Bogot\'a has reduced the mean travel
time in a $40\%$, accidents in an $80\%$ and pollution in a $50\%$. However,
there are still many improvements to be done in the future.

During the last 10 years, cellular automata models (CA) have been applied with
success to traffic simulations.  These models are able to reproduce the
macroscopic properties of highway traffic from the microscopic behavior of
each car. The first model (STCA) \cite{nagel,nagel1}, proposed by
Nagel and Schrekenberg in 1992, is just based in the distance to the next car
($gap$$=$$\Delta x$) and the maximal speed, but leads to a quite realistic
flow-density relation (fundamental diagram) and reproduces well the
spontaneous jam and shock waves formations in highways \cite{nagel1,20}. Later developments add other
elements to STCA, like improved gaps including the speed difference to the car
ahead \cite{wolf,Barrett}, the speed at the previous time step
\cite{wolf,Barrett}, and many other parameters. Some theoretical and practical
studies have extended these models to two- and three-lane highways
\cite{simon}. 

The drivers' driving is very different from city to city, and a
realistic traffic model should keep in mind the particularities of each
place. Simulations and studies of this type have been carried out in cities
like Portland, Los Angeles, Tokyo \cite{wagner,wagner1,libro}.  However, there
are not such studies performed in Colombia, even in Bogot\'a.

In this work we propose a car cellular automaton model that reproduces the
experimental behavior of traffic flows in Bogot\'a. Our model includes three
elements. The first one are the gaps the driver uses to decide to
brake (brake gap $gap_{brake}$) or accelerate (acceleration gap
$gap_{accel}$). They are, in general, different (hysteresis) and both depend on
the speed. 
The second element is the time it takes the car to reach the next discrete
speed value (retarded acceleration, $t_{up}$). 
The last one is an instantaneous brake reaction (that) we have observed when the car ahead brakes.
The parameters of our model were obtained from direct
measurements inside a car that was running on Bogot\'a's highways. With these parameters,
simulations were performed to construct the flow-density fundamental diagram.
This result was compared with experimental measures from Bogot\'a's
highways. 
The model was also used to compute shock waves, both in the free and
congested regimes.
Finally, we studied the phase transition between free and congested traffic
(jamming transition) by computing both the relaxation time and the order
parameter according to their definition in \cite{kertsé,kertse1}. This last
study was performed on both the deterministic model proposed above and on the
same model plus a probabilistic spontaneous-brake rule, as in STCA. 

The paper proceeds as follows. In section 2, we make a detailed description of
our model, the values of the measured parameters and the vehicle
rules. Section 3 we show the simulation results and compare them
with experimental data. Section 4 includes our study on the jamming
transition. Section 5 contains the main conclusions and discussions of our
work. Finally, appendix A and B describe in detail the experimental methods we
used to obtain the model parameters and the experimental flux-density
diagrams.
\pagebreak

\section{Our Model Description}
In our model, the highway is represented by a one-dimensional array of length
L with periodic boundary conditions. Each site of the array is a cell of
length $2.5m$, that is a finer discretization than the used in STCA
model. Vehicles can only have integer velocity values,
$v$$=$$0,1,...,v_{\max}$. We used $v_{\max}$$=$$7$ and a speed unity
$v_{unity}$$=$$10 \frac{km}{h}$.  
This corresponds to time steps of $t_{step}$$=$$0.9s$, that is near to the
usual value to driver's reaction time. These discretizations are usual for any
CA traffic model, with the only difference that a vehicle occupies two
consecutive cells, the length of a car ($4.5 m$) plus the distance between
cars in a jam ($1 m$). Then, the maximal number of vehicles in the highway is
given by $N$$=$$\frac{L}{2}$.
At time $t$ the n-vehicle is completely defined by: its position
$x_n(t)$, its velocity $v_n(t)$ and its brake-light status,
$b_n(t)$, which is $b_n$$=$$1$($0$) when the driver brakes (or not) at the
previous time step ($t-1$), like \cite{toward}. The effective $gap$
is defined as $gap$$=$$\Delta x(t)+ \Delta v(t)$, where $\Delta
x(t)$$=$$x_{n+1}(t)-x_n(t)-1$ is the number of cells empty to the vehicle
ahead and $\Delta v(t)$$=$$v_{n+1}(t)-v_{n}(t)$ is the speed difference to the
car ahead.

As already mentioned, our model includes three elements: the hysteresis
between brake and accelerate gaps, the retarded acceleration and the
instantaneous break. The three parameters $gap_{brake}$, $gap_{accel}$ and
$t_{up}$, are function of speed and they represent on the whole the drivers'
driving. These parameters were experimentally found from inside a car on
Bogot\'a's highways (see appendix A) and are summarized in table \ref{tablas}.

\begin{table}[htb]
  \centering
  \begin{tabular}{|c|c|c|c|}\hline
    \multicolumn{4}{|c|}{Drivers' driving in Bogot\'a}\\
    \hline \hline
    \textit{Speed}&\textit{$gap_{brake}$}&\textit{$gap_{accel}$}&\textit{$t_{up}$}\\
      \hline 0&0&3&1\\
      \hline 1&3&4&1\\
      \hline 2&3&5&1\\
      \hline 3&4&5&1\\
      \hline 4&5&6&2\\
      \hline 5&6&7&2\\
      \hline 6&6&8&2\\
      \hline 7&7&9&2\\ \hline  \end{tabular}
  \tcaption{{\footnotesize Drivers' driving from Bogot\'a.}\label{tablas}}
\end{table}
Summarizing, all cars execute in parallel the following set of rules:
\pagebreak

\centerline{\bf{Rules}}

\begin{itemize}
\item Compute its $gap(t)$.
\item Read its parameters $gap_{accel}$, $gap_{brake}$ and $t_{up}$ from table \ref{tablas}.
  \begin{itemize}
  \item \textit{normal brake}: If $gap(t)$$\le$$gap_{brake}$, speed down to the maximal speed $v(t+1)$
  such that $gap'_{brake}$$\leq$$gap(t)$$\leq$$gap'_{accel}$, where
  $gap'_{brake}$ and $gap'_{accel}$ are the parameters at speed $v(t+1)$. In
  addition, let $delay$$=$$0$ and turn on brake lights ($b_n(t+1)$$=$$1$).
  \item If $gap$$\geq$$gap_{accel}$, then
    \begin{itemize}
    \item \textit{instantaneous brake}: If $gap(t)$$\leq$$gap_{accel}+2$
      and the brake lights of the car ahead are on ($b_{n+1}(t)$$=$$1$), let $v(t+1)$$=$$v(t)-1$ (brake), turn on brake
      lights ($b_n(t+1)$$=$$1$) and let $delay$$=$$0$
      .
    \item \textit{accelerate}: Else, turn off brake lights ($b_n(t+1)$$=$$0$) and
      \begin{itemize}
      \item If $delay$$=$$=$$t_{up}$, let $v(t+1)$$=$$v(t)+1$ (accelerate) and
      let $delay$$=$$0$.
      \item Else, let $delay$$=$$delay+1$ and preserve $v(t+1)$$=$$v(t)$.
      \end{itemize}
    \end{itemize}
  \item Otherwise, let $delay$$=$$0$, turn off brake lights ($b_n(t+1)$$=$$0$)
  and preserve $v(t+1)$$=$$v(t)$. 
  \end{itemize}
\item Finally, move $v$ cells ahead,
\begin{eqnarray}
x(t+1)= x(t)+v(t+1)\quad.
\end{eqnarray}
\end {itemize}

The counter $delay$ defines if $t_{up}$ has been completed. The variable
$b_{n+1}(t)$ defines the brake light status of the car ahead. The
\textit{instantaneous brake} rule represents the braking reaction we have
observed when the car ahead also brakes. This reaction is observed for all
distances but is just included in the gaps when $gap\le gap_{accel}$. Thus, we
have included it as an additional rule only it $gap_{accel}\leq gap (t)\leq
gap_{accel}+2$ through a brake light on each car.

\section{Results}

The system starts with a initial configuration of $N$ cars, with 
random distributions of speeds and positions and $b_j(0)$$=$$0$ \textit{for
  all j}. In order to prevent traffic accidents, as a previous step, we limit
the speed values to the headway ($v_{ini}$$\le$$\Delta x$). Starting from this
initial configuration, we measure the average velocity $\bar{v}(t)$ over all
cars at each time step, $t$. After many time steps ($t$$\to$$\infty$), when
the system reaches a stationary velocity state $\bar{v}(t)$$=$$v(\infty)$, the
flow is computed by the relation
$q$$=$$v(\infty)\cdot\rho$$=$$v(\infty)\cdot\frac{N}{L}$. The whole process
is repeated $100$ times for each density value, just to make statistics, 
and, so, the fundamental diagram is obtained.

\begin{figure}[!htp]
\centering
\includegraphics[width=2in]{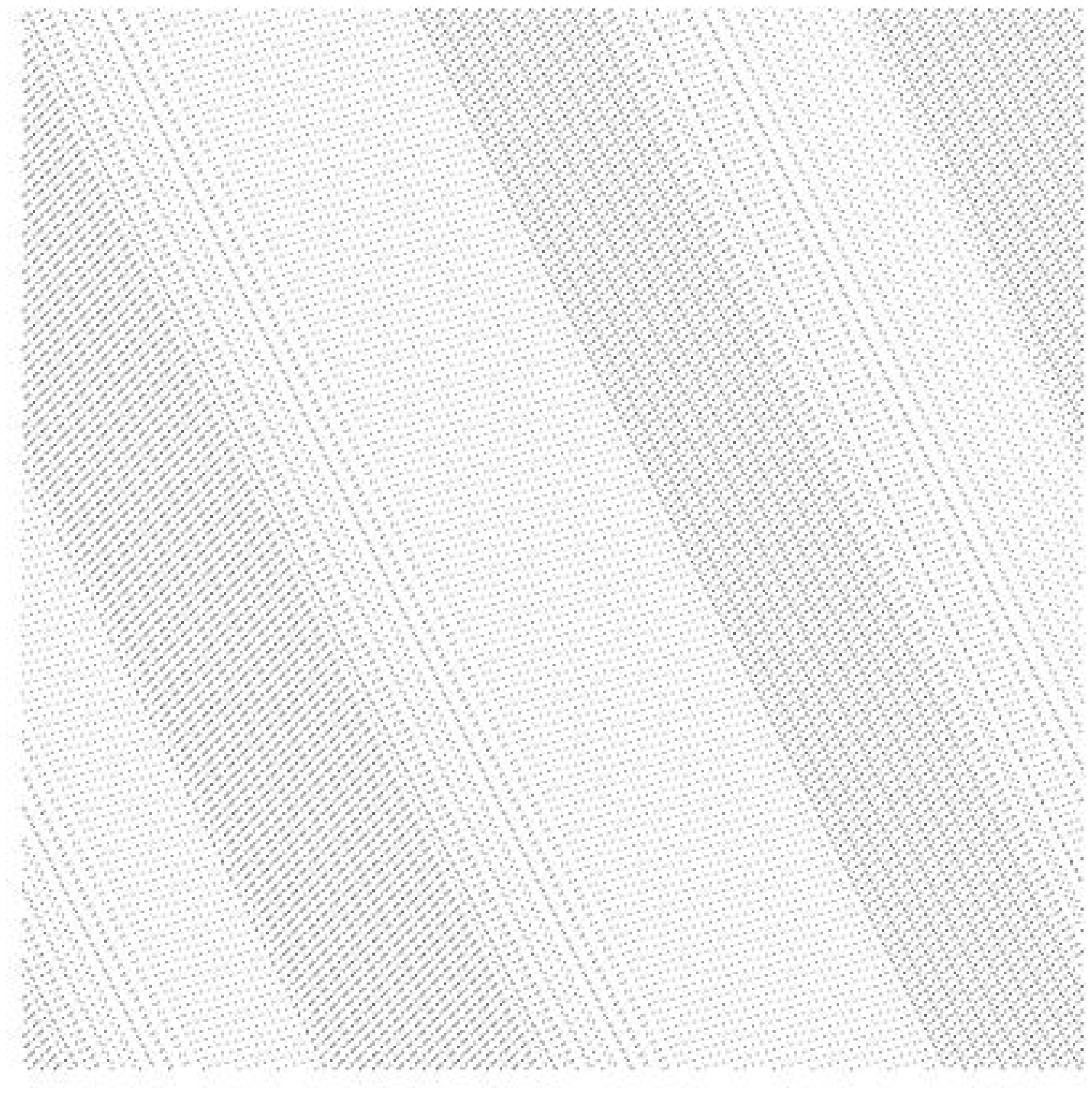}\hspace{1cm}
\includegraphics[width=2in]{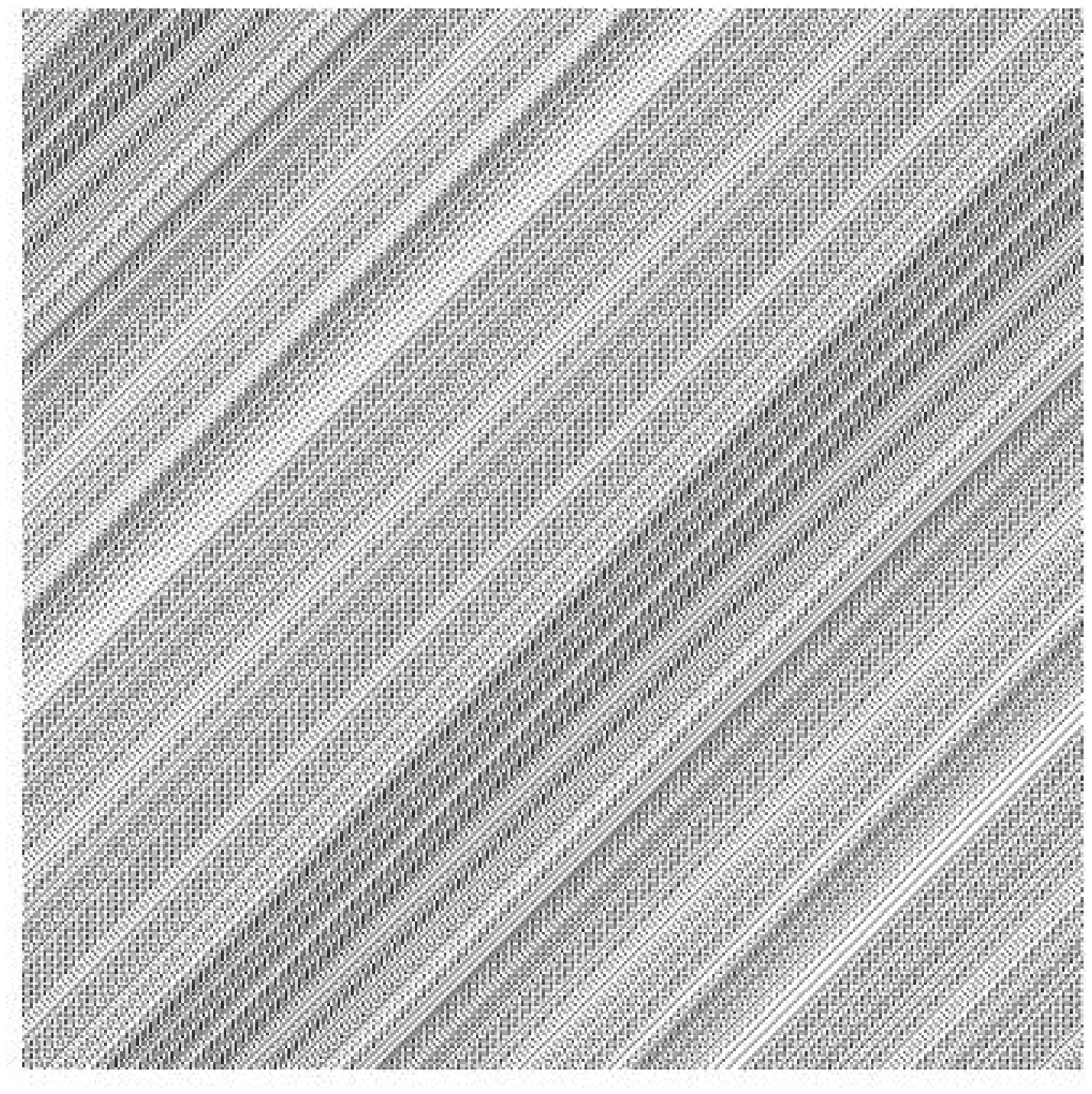}\\
a)\hspace{5.5cm} b)
\fcaption{{\footnotesize Space-time plot for a road of $L$$=$$1000$: a)
    $\rho$$=$$0.16$ ($\rho\leq\rho_c$), b) $\rho$$=$$0.6$
    ($\rho\geq\rho_c$). Cars move from left to right and time runs from
    top to bottom. One can observe the schock waves and jams
    formations. If $rho\le\rho(q_{\max})$($rho\ge\rho(q_{\max})$), these
    formations are moving forward (backward)}\label{jams}}
\end{figure}

Our model is able to reproduces the phases observed in real traffic: free-flow,
synchronized, and stop-and-go. The shock waves and jams formations for these
flow regimes can be observed in space-time plots (figure \ref{jams}). 

\pagebreak

\begin{figure}[!htp]
  \centering
  \includegraphics[width=3.5in]{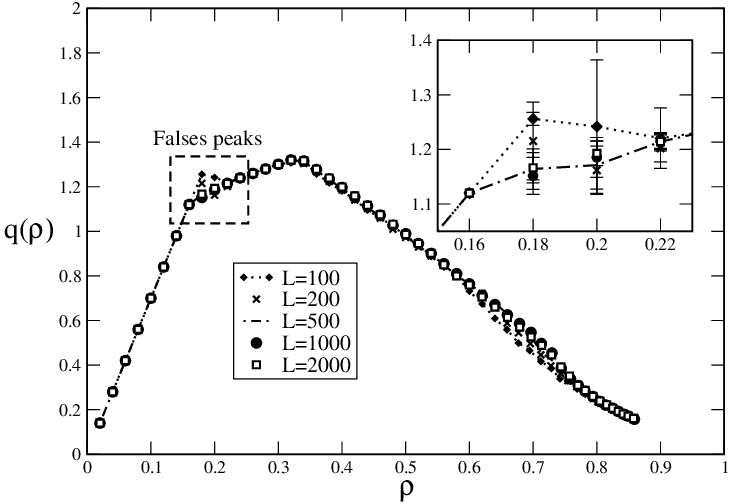}
  \fcaption{{\footnotesize Effects of small system sizes ($L$) into the
  simulated fundamental diagram. One observes the false peak of maximal flux
  due to the finite boundary conditions and how it disappears when
  $L$$=$$2000$ }\label{efectos}}
\end{figure}

Before comparing the simulation results
with the experimental data, we study in detail the finite size effects.
For small systems, appears in the fundamental diagram a false peak of maximal
flow, which disappears when the system is sufficiently large. Figure
\ref{efectos} shows the fundamental diagram for different system sizes. One
can observe that 
the false peak appears around $\rho$$=$$0.12$ (in the synchronized traffic
phase), and it has completely disappeared for $L$$=$$2000$. These
false peaks have also been observed in many other models
\cite{krauss1,krauss}. In \cite{krauss} they are interpreted as the
coexistence of two phases of traffic flow in a dynamic equilibrium. In
contrast, we have found that these false peaks are due to extraordinary
configurations that exists only in small systems. Due to the periodic-boundary
conditions, it is possible to think a configuration where all cars have the
maximal speed and all have the same gap, equals to that maximal speed. 

In these configurations the system does not relax, but remains forever, 
with a mean velocity that is larger than the average velocity  for
the relaxed system. Thus, they push up the values of $\langle
v(\infty)\rangle$, where the $\langle \rangle$ denotes the average on many
realizations, generating the false peak. In conclusion, the peak should be 
located at densities $\rho$$=$$1/v_{\rm max}$$=$$0.14$, in good agreement with
the computational results.

Since our model takes into account three elements, we were interested in
looking at the effect of including each one of them, one by one, namely:
the hysteresis between brake and acceleration gaps, the retarded acceleration
and the instantaneous brake.

\begin{figure}[!htp]
  \centering
  \includegraphics[width=3.5in]{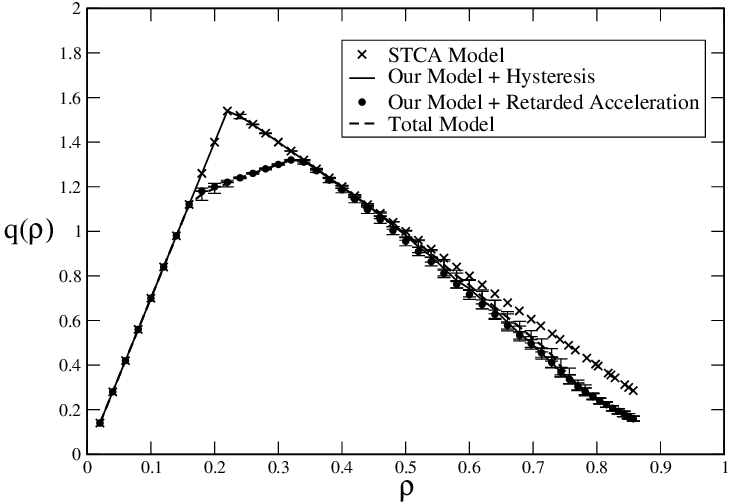}
  \fcaption{{\footnotesize Effects that have each one of the elements of our
  model on the shape of the fundamental diagram. Also is shown the fundamental
  diagram of the STCA model with our time and velocity
  discretization.}\label{partes}}
\end{figure}

Figure \ref{partes} shows the fundamental diagrams obtained from our model
with (a) brake and acceleration gaps only, (b) gaps plus retarded acceleration
and (c) all three elements. They are compared with (d) 
the fundamental diagram from the STCA model with our time and
space discretization. One notes small differences between the fundamental
diagrams corresponding to STCA and the most elementary model. This means that
the hysteresis doesn't have any relevant effect on the fundamental diagram. Its
consequences, however, should be seen in the behavior of jams and shock waves,
but they will be not investigated here and they would be theme of future work.

When the retarded acceleration is included, two effects emerge, namely: a
decrease of the maximum flow and a its shift towards larger
densities. Finally, the \textit{instantaneous brake} effect shows up slightly
into the congestion region.
\vspace{1cm}
\pagebreak

\begin{figure}[!htp]
\centering
\includegraphics[width=3.5in]{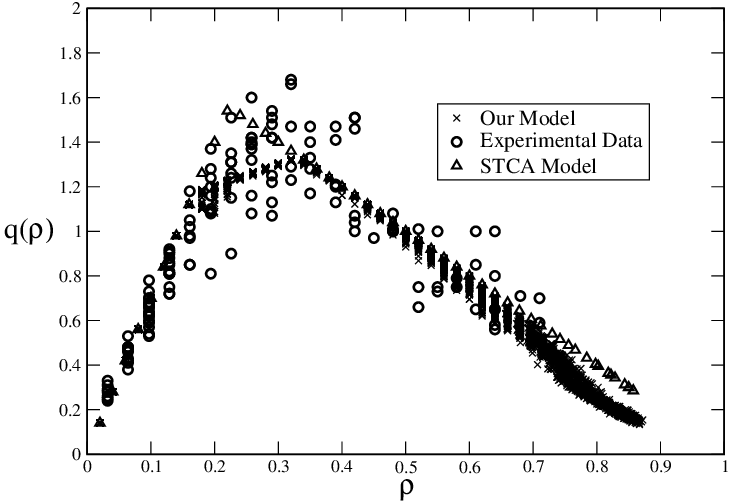}\vspace{1.3cm}
\includegraphics[width=3.5in]{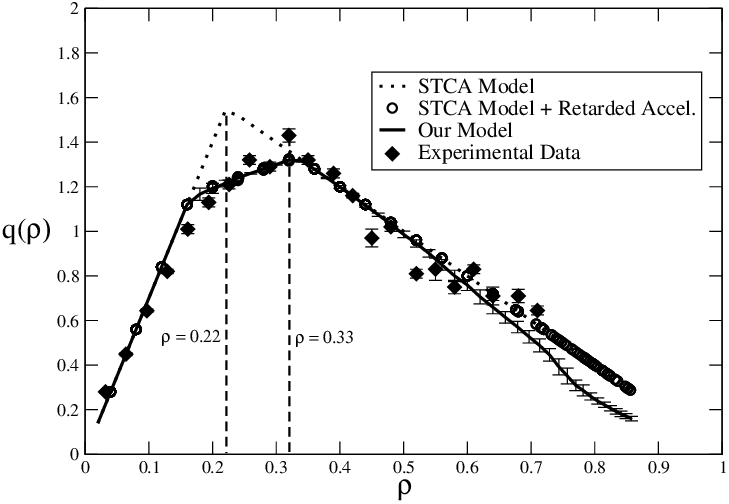}
\fcaption{{\footnotesize Comparison of the flow-density diagram for the
    simulated models with the experimental measures. Both simulated models,
    our model and de STCA model plus retarded acceleration, are in good
    agreement with the actual data. Above: dispersed data. Below: averaged
    data.
}\label{diaexp}}
\end{figure}

To validate our model we performed a comparison with experimental measurements.
Figure \ref{diaexp} compares the flow-density diagram from
our model with measurements over Bogot\'a's highways on broad density ranges,
both as  a) dispersed data and b) averaged data. A second alternative,
consisting of a STCA model plus retarded acceleration , is also included.  

On one hand, one observes that both
simulations are in good agreement with the experimental data, especially if
we compare the numerical values, simulated and measured, of the road capacity
and the density of maximal flux. 
\begin{eqnarray}
\rho(q_{\max})_{exp}=0.33(3)\quad  \quad q_{\max exp}=1.43(6),\\
\rho(q_{\max})_{sim}=0.33(4)\quad \quad q_{\max sim}=1.320(4).
\end{eqnarray}
The difference in the scatter between simulated and experimental data shows up
(at least in part) because the first ones are averages over much more cells than the second
ones. The two models have different behaviours just
in the zone of high congestion, were our model predicts lower fluxes. This is
the effect of hysteresis that is expected to play a role in jam formations. 
On the other hand, these values of road capacity are slightly larger than those
measured in other countries \cite{libro}. This suggests
that the traditional Bogot\'a's aggressive driving makes the traffic flux more
efficient. The price is, however, one of the most high rates of fatal victims
on the world \cite{fatal} (in fact, one out of six victims of violent causes in
Colombia dies in a car accident \cite{fatales}). For a recent work including
agressive drivers see \cite{lee04}.

Cellular automata models for traffic flow exhibits sometimes a phase 
transition from a free-flow phase to a congested phase
\cite{kertsé,kertse1,kertse2}. In most cases this 
transition is first order, but some models, like STCA, shows a second-order
phase transition at a single point in the phase space \cite{Jost}. 
The relaxation time and the order parameter are the
quantities which characterize this transition.

To compute the relaxation time we employed 
the definition of Cs\'anyi and Kert\'esz \cite{kertse1}. As they do, we start from
a configuration random initial positions and zero speed for all cars. Then,
the relaxation time is computed as 

\begin{equation}
\tau=\int_{0}^{\infty}[\min \{v^*(t),\langle \bar{v}(\infty)\rangle \}-\langle
\bar{v}(t) \rangle]dt \quad ,
\end{equation}\label{tau}
where $v^*(t)$$=$$t$ denotes the speed a car obtains at time $t$ when there
are no cars ahead (free acceleration).

\begin{figure}[!htp]
\centering
\includegraphics[width=3.5in]{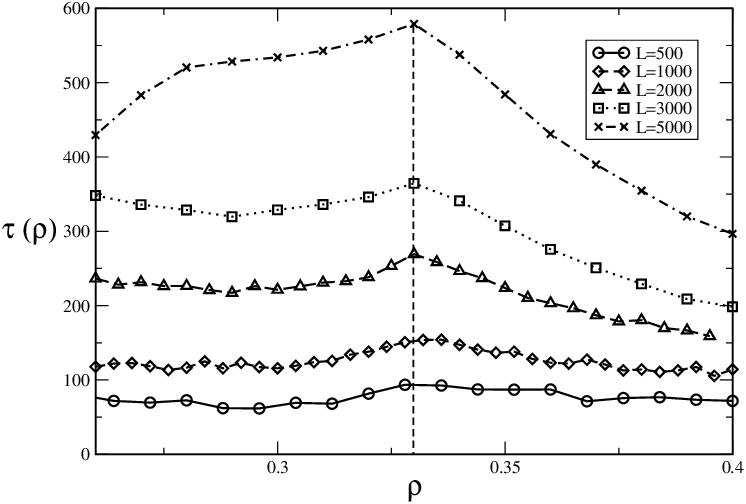}
\fcaption{\footnotesize Relaxation time $\tau$ near the transition density
  $\rho_c$ for different system sizes.}\label{tiempos}
\end{figure}

As we already know, a characteristic feature of a second order phase
transition is the divergence of the relaxation time at the transition
point.
Figure \ref{tiempos} shows how the relaxation time has a maximum at a
pseudo-critical density $\rho_c$$=$$0.33$, which is the 
same value of maximal-flow density one can read from the flux-density
fundamental diagram (figure \ref{diaexp}), as expected for a deterministic
model \cite{kertsé}. 

The typical order parameter for traffic cellular automata is \cite{kertsé}.
\begin{equation}
  m=\frac{1}{L}\sum_{i=1}^{L} n_i n_{i+1}\quad.
\end{equation}\label{mu}
Since a car in our model occupies two consecutive cells, we redefine the order
parameter as
\begin{equation}
  m=\frac{1}{L}\sum_{i=1}^{L} [n_i n_{i+2}+ n_i (1-n_{i+2})n_{i+3}]\quad.
\end{equation}\label{mu2} 

\begin{figure}[!htp]
\centering
\includegraphics[width=3.5in]{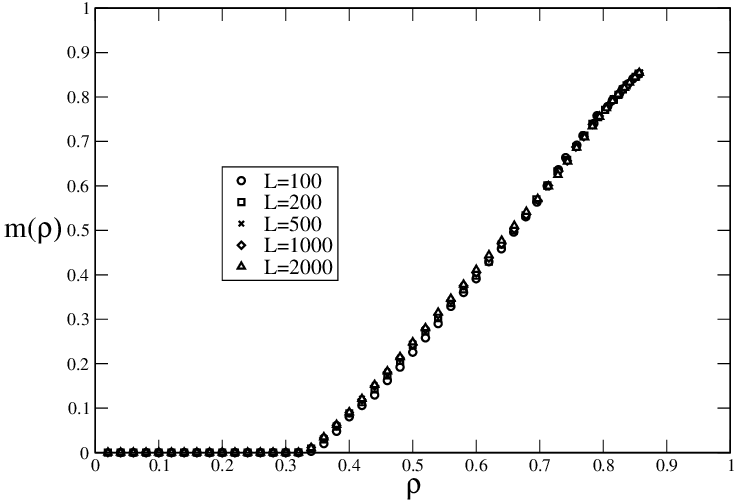}
\fcaption{{\footnotesize Order parameter for different system sizes. Below the
    transition density $m(\rho)$ decreases to zero.}\label{parametro}}
\end{figure}

Like \cite{kertsé}, figure \ref{parametro} shows that the order parameter
exhibits a continous transition. The situation is quite similar to the
behavior of the order parameter in finite systems. That is, the order
parameter have small values for small density, and around the $\rho_c$ it
begins to have a non-zero value.
\vspace{1cm}

Finally, we wanted to explore what occurs when we add the randomization rule
of the STCA model. This is,

\begin{itemize}
\item Randomization: If after the above steps the velocity is larger than zero
  ($v\ge0$), then, with probability p, let $v(t)$$=$$v(t)-1$.
\end{itemize}

\begin{figure}[!htp]
\centering
\includegraphics[width=3.5in]{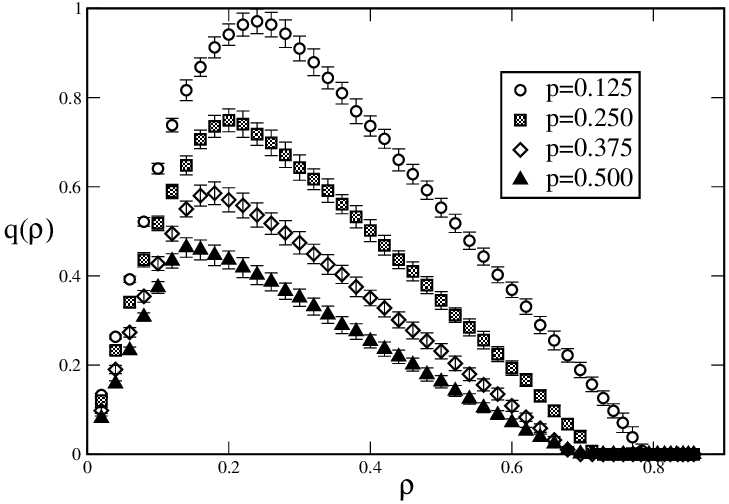}
\fcaption{{\footnotesize Fundamental diagram for different values of $p$. This
    figure corresponds to our model plus randomization rule of the STCA model.
}\label{totalSTCA}}
\end{figure}

Figure \ref{totalSTCA} shows the fundamental diagrams for differents values of
$p$.
One observes that when the noise p increases, the flow decreases and the
maximum towards smaller densities. In addition, for larger densities, the
randomization rule causes an early collapse in the system.
\vspace{1cm}

\begin{figure}[!htp]
  \centering
  \includegraphics[width=2.4in]{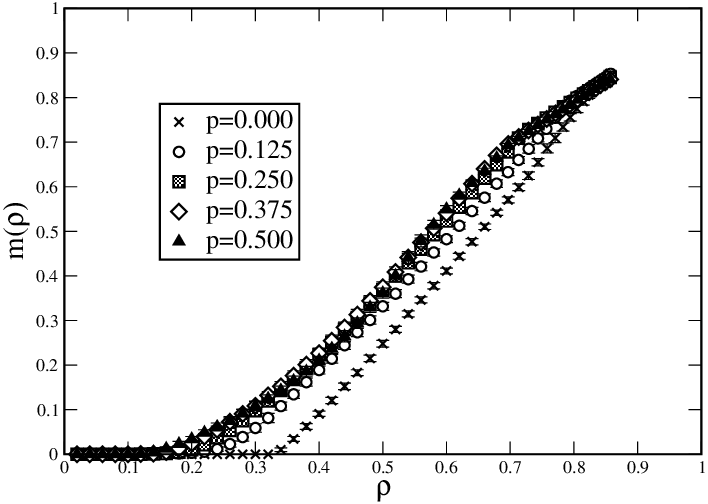}\quad
  \includegraphics[width=2.4in]{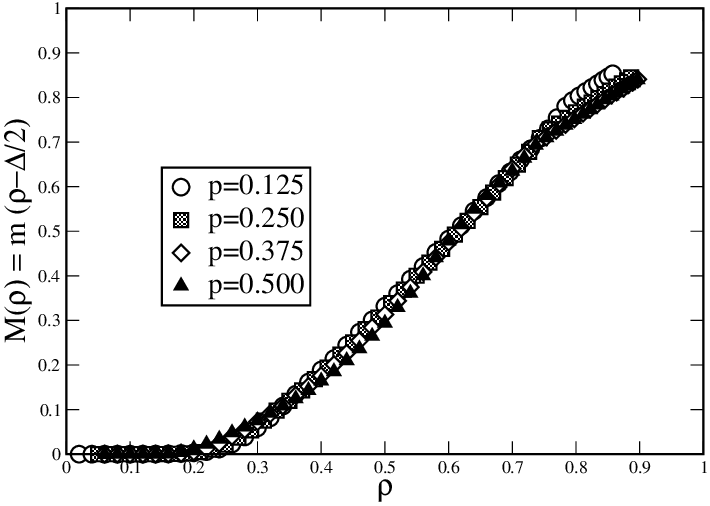}\\
  \centerline{a)\hspace{6cm}b)}
  \fcaption{{\footnotesize  Behavior of the order parameter for Our model with
  a randomization rule. a) Order parameter for different values of $p$ and
  b) Scaling plot for the order parameter (excluding
  $p$$=$$0$).}\label{paramSTCA}}
\end{figure}

Figure \ref{paramSTCA} shows the order parameter behavior for differents
values of $p$. It converges smoothly to zero for densities smaller to
$\rho_c$. To the right hand, we performed the respective scaling. The
following form summarize the simple scaling:

\begin{equation}
M(\rho)=m(\rho+\frac{\Delta}{2})\quad.
\end{equation}

Herein $\Delta$ is the shift of the transition density compared to the same
value for $p$$=$$0.125$, because it was impossible to make the scaling for the
deterministic case.

\section{Conclusions}  

In this work, we present one of the first cellular automaton models that
applies to the reality of Bogot\'a. Their objective is not just to propose a
first model, but  also, to investigate which models apply to our city. The
work includes  also measures of the driving parameters of a car in Bogot\'a
and of the fundamental diagram of a highway. 

Our model includes three elements, to summarize are the drivers' driving:
hysteresis between brake and acceleration gaps, retarded acceleration and
instantaneous brake.

The simulation results are in a good agreement with the experimental
measurements.  That is, the simulated fundamental diagram reproduces
successfully the shape of the fundamental diagram and the numerical values for
both the road capacity ($q_{\max}$) and the density of maximal flux
($\rho(q_{\max})$).  The obtained values are:

\begin{eqnarray}
\rho(q_{\max})_{exp}=0.33(3)\quad  \quad q_{\max exp}=1.43(6)\\
\rho(q_{\max})_{sim}=0.33(4)\quad  \quad q_{\max sim}=1.320(4).
\end{eqnarray}

For small sistem sizes we have found that the false peak of maximal flux  in
the fundamental diagram is due to some extraordinary configurations. In these
configurations the system remains in a not-relaxation state, with a mean
velocity larger than $\langle v(\infty)\rangle$. Finally, these extraordinary
configurations only exist for systems with $L\le2000$. 

The hysteresis element in our model doesn't have any relevant effect on the
fundamental diagram. If we compares it with the STCA model, the differences
appears only into the congestion region. However, we believe that the
spontaneous formations of jams and shock waves should be affceted by this
elements, and this is an interesting area of future work. In contrast, the
retarded acceleration is much more important. It fixes the 
real shape of the fundamental diagram and the characteristic values of
$q_{\max}$ and $\rho(q_{\max})$ . 
Moreover, an STCA model plus the retarded acceleration is enough to reproduce
the fundamental diagram we found in Bogot\'a's highways.

By looking at the phase-transition, we found qualitatively the same behavior
of both relaxation time and order parameter as shown in
\cite{kertsé,kertse1}. One observes a maximum of the relaxation time $\tau$
near the density transition $\rho_c$, specially for the deterministic case
$\rho_c$$=$$\rho(q_{\max})$. The divergence is more marked when the system size
increases. The behavior of the order parameter shows a smooth convergence to
zero for small densities.

Summarizing, both our model an the STCA model plus retarded acceleration
reproduces the fundamental diagram for Bogot\'a. They coul be used to perform
more complicated simulations in future works, like semaphorized, intersections
and even the Bogot\'a's net of main highways. It also remains to study the jam
formation in both models.

\section{Acknowledgments}  
We thank M. Schreckenberg and an annonymous referee for valuable
comments, corrections and suggestions. 

\appendix
{
: Our Model Parameters}
{
By installing a videocamera inside a car, we taped the driving of a driver on
the highways in Bogot\'a. The videocamera was calibrated with the plates of
the other cars, therefore  we made a chart, which relates the longitude of the
plates with the distance to the car in front, $\Delta x$. During the trips,
the copilot registers the car speed the velocity and if the driver brakes,
accelerates or preserves this speed. With the films on hand, one can know at
any time the vehicle speed, its state (braking, acceleratimg or preserving)
and the distance to the car in front ($\Delta x$). The relative velocity
$\Delta v$ is calculated as
\begin{equation}
\Delta v=\frac{\Delta x_{2}-\Delta x_{1}}{2},
\end{equation}
where $\Delta x_{1}$ is the usual $\Delta x$, and $\Delta x_{2}$ is the
distance to the car in front measure after two seconds.

To include the retarded acceleration it is neccesary to measure the parameter
$t_{up}$. For this purpose, the car was accelerated from $0\frac{km}{h}$ to
$100\frac{km}{h}$ and we measure the time, it takes to reach the next discrete
speed value, in steps of ($10\frac{km}{h}$).

\begin{figure}[!htp]
  \centering
  \includegraphics[width=3in]{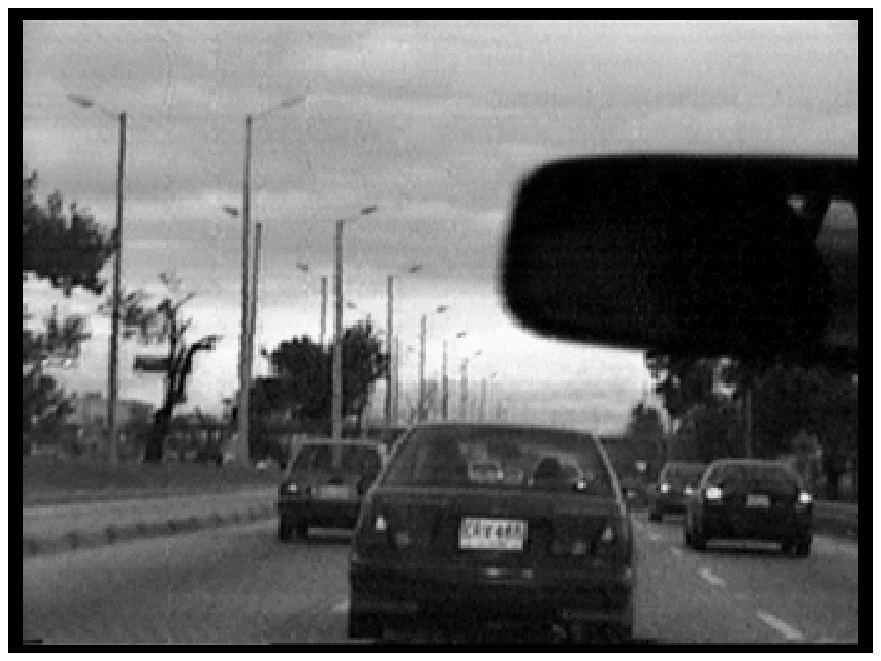}
  \fcaption{{\footnotesize Digitized pictures of the tape when the parameters
  of the model were measured.}\label{parametros1}}
\end{figure}
}

\appendix
{
: Experimental Measurements}
{
The experimental data were obtained by capturating on tape (from a pedestrian
bridge), the traffic flow in a highway segment. The $30^{th}$ avenue is
perhaps the most important avenue of Bogot\'a,  because it communicates
directly the south with the north of the city. At daily rush hours the
$30^{th}$ avenue is not able to cope with the demand and bored traffic jams
are generated. We chose for our measurements the south-north high-speed lane
of the $30^{th}$ avenue between $53^{th}$ street and the Camp\'in Football
Stadium.

This is a two-lane sector and have a length of $L=$$169(5)m$$\approx$$62(2)$
cells, i.e, a maximal number of cars $N_{\max}$$=$$31(1)$. 

We carried out the following process:

\begin{itemize}
\item At each time, the density of the system is computed by
  $\rho$$=$$\frac{N(t)}{N_{\max}}$, where N(t) is the number of cars over the
  highway sector.

\item The velocity of the system is the average velocity over the N cars. It
  is calculted as an aritmethic mean, $\bar{v}(t)$$=$$\frac{1}{N} \sum_{j=1}^N
  \bar{v_j}(t)$, where $\bar{v_j}(t)$ is the average velocity of each car over
  the highway sector.

\item Finally the flow is calculated as $q$$=$$\rho \cdot \bar{v}(t)$.

\end{itemize} 

With this process we obtained the experimental data to perform the
flux-density fundamental diagram.

\begin{figure}[!htp]
  \centering
  \includegraphics[width=3in]{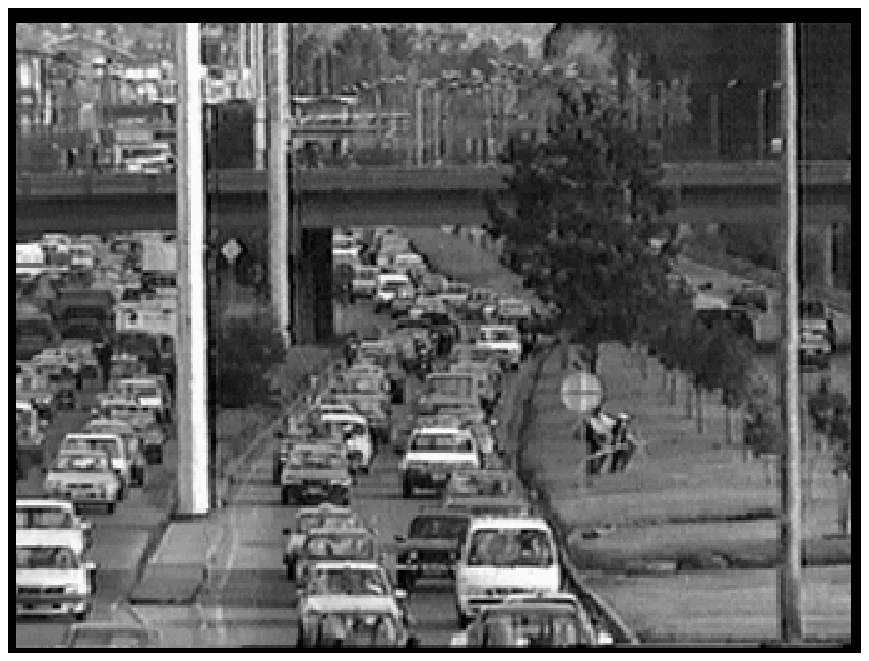}
  \fcaption{{\footnotesize  Digitized picture of the tape, when the
  experimental fundamental diagram was measured.}\label{parametros}}
\end{figure}
}

\nonumsection{References}

\end{document}